\documentclass[12pt]{iopart}
\usepackage{iopams}

\usepackage{color}

\usepackage[normalem]{ulem}

\usepackage[hidelinks]{hyperref}

\usepackage[utf8]{inputenc}
\usepackage{nicefrac}
\usepackage{upgreek}
\usepackage{braket}
\usepackage{cite}

\usepackage{graphicx}
\usepackage{graphics}

\newcommand{\f}{\theta}
\renewcommand\sout{\bgroup \color{blue} \ULdepth=-.5ex \ULset}

\begin{document}

\title{Net-proton number fluctuations in the presence of the QCD critical point}

\author{Micha\l~ Szymański$^1$,
        Marcus ~Bluhm$^{1,2}$,
        Krzysztof  Redlich$^{1,3}$,
        Chihiro  Sasaki$^1$
        }

\address{$^1$ Institute of Theoretical Physics, University of Wroclaw,PL-50204 Wroc\l aw, Poland}
\address{$^2$ ~SUBATECH UMR 6457 (IMT Atlantique, Universit\'e de Nantes, IN2P3/CNRS), 4 rue Alfred Kastler, 44307 Nantes, France}
\address{$^3$ Extreme Matter Institute EMMI, GSI, Planckstr. 1, D-64291 Darmstadt, Germany}
\ead{michal.szymanski@ift.uni.wroc.pl}
\vspace{10pt}
\begin{indented}
\item[]May 2019
\end{indented}

\begin{abstract}
Event-by-event fluctuations of the net-proton number studied in heavy-ion collisions provide an important means in the search for the conjectured critical {end} point (CP) in the QCD phase diagram. We propose a phenomenological model in which the fluctuations of the chiral critical mode couple to protons and anti-protons. This allows us to study the behavior of the net-proton number fluctuations in the presence of the CP. Calculating the net-proton number cumulants, $C_n$ {with n=1,2,3,4}, along the phenomenological freeze-out line we show that  {the ratio of variance and mean $C_2/C_1$, as well as kurtosis $C_4/C_2$}  resemble qualitative properties observed in data in heavy-ion collisions as a function of beam energy obtained by the STAR Collaboration at RHIC. In particular, the non-monotonic structure of the {kurtosis} and smooth change of the $C_2/C_1$  ratio with beam energy could be   due to the CP located near the freeze-out line. {The skewness, however, exhibits properties that are in contrast  to  the criticality expected due to the CP.}  The dependence of our results on the model parameters and the proximity of the chemical freeze-out to the critical point are  also discussed.
\end{abstract}

%
%
%
%
%

{Delineating  the phase diagram of quantum chromodynamics
(QCD) at finite temperature $T$  and  baryon-chemical potential $\mu_B$
is a challenging problem in theoretical and experimental  studies \cite{L1,L2,L3,L4,Lm1,Lm2,Lm3,Lm4,NPBM,T1,T2,Luo:2017faz,Ex1}.  The Beam Energy Scan (BES) at
the Relativistic Heavy Ion Collider (RHIC)}  \cite{Luo:2017faz,L2016}  {has been
dedicated to the search for the conjectured QCD critical end
point (CP)  through
the systematic studies of various observables.}
In particular, a non-monotonic behavior of {the
fluctuations of conserved charges}  {with}  beam energy ($\sqrt{s}$) is {considered as a conceivable experimental   signature of the chiral critical behavior and of the CP in heavy-ion collisions \cite{T2,Stephanov:1998dy,Stephanov:1999zu,Asakawa:2015ybt,T3,T4,T5}.}  {Such a behavior is}   typically associated with the divergence of the correlation length and the fluctuations of the critical mode ($\sigma$) at the {CP}  \cite{Stephanov:1998dy,Stephanov:1999zu,Cp1,Cp2,Cp3,Cp4,Cp5}. Another, indirect way of verifying the existence of the CP is given by searching for non-uniform structures in multiplicity distributions due to domain formations in {the region of } the first-order phase transition  adjacent to the critical point
\cite{Steinheimer:2012gc,Herold:2013qda,S1}.

 Measurements of event-by-event fluctuations in the net-proton number~\cite{Adamczyk:2013dal,Luo:2015ewa,L2016,Thader:2016gpa}, as a proxy for net-baryon number, the net-electric charge~\cite{Adamczyk:2014fia} and the net-kaon number~\cite{Adamczyk:2017wsl}, as a proxy for net-strangeness, have been performed in heavy-ion collisions at RHIC and LHC energies. While non-monotonic structures  in the still preliminary STAR Collaboration data on  higher-order net-proton fluctuations were indeed observed \cite{Luo:2015ewa,Luo:2015doi,Thader:2016gpa}, {their} unambiguous interpretation as the consequence of the presence of the CP  has not  been achieved yet \cite{T4,T5,V1,V2}.

{Different QCD-like} effective models indicate,  that the chiral CP  belongs to the {Z(2)} static universality class of the $3$-dimensional Ising model,  and that the chiral critical mode can be associated with the order parameter, the chiral condensate~\cite{Cp3,Cp4,Cp5,Berges:1998rc}. For  a static and infinite system in thermodynamic equilibrium, the scaling of the fluctuations of $\sigma$  modes with the diverging correlation length at the critical point is governed by the Z(2) critical exponents~\cite{Stephanov:2008qz}. However, a medium created in a heavy-ion collision is neither static nor infinite and due to its expansion dynamics,  the  non-equilibrium effects can  play an  important  role. The temporal growth of the correlation length  is already dynamically limited by the phenomenon of critical slowing down~\cite{Berdnikov:1999ph,Son:2004iv}. {In addition,} non-equilibrium effects can lead to the retardation and damping of the critical signals~\cite{Nahrgang:2011mg,Herold:2013bi,Mukherjee:2015swa,Herold:2016uvv}. {Furthermore,} even in thermodynamic equilibrium, the exact charge conservation~\cite{C1,C2,C3}, { volume fluctuations}~\cite{Vol1,Schuster:2009jv,C2,Nahrgang:2018afz}, {and further sources of non-critical fluctuations in the data}~\cite{Nahrgang:2016ayr}, {as well as, late hadronic stage processes}~\cite{Jiang:2015hri,Kitazawa:2011wh,Kitazawa:2012at,Nahrgang:2014fza,Bluhm:2016byc}, can modify signals of the critical fluctuations.

Given the above mentioned challenges it becomes clear that the theoretical description and  interpretation of  data on fluctuation observables  require special care and   eventually a dynamical framework to match the conditions expected in heavy-ion collisions. Nevertheless,   numerical simulations of the critical dynamics require also input from static, equilibrium models to provide an analytic benchmark~\cite{Herold:2016uvv,Nahrgang:2018afz,Herold:2017day,Bluhm:2018qkf}. One of  such  phenomenological models,  that accounts for the critical fluctuations in heavy-ion collisions has been formulated in Ref.~\cite{Bluhm:2016byc}.  There,  the fluctuations of the critical mode were  {linked}  to (anti-)protons and resonances by allowing the particle masses to fluctuate as a consequence of the (partial) mass generation through the coupling to $\sigma$. As was demonstrated in Ref.~\cite{Bluhm:2016byc}, the non-monotonicity observed  in the  higher-order fluctuation data of STAR Collaboration~\cite{Luo:2015ewa,Luo:2015doi,Thader:2016gpa} can be qualitatively   described within  such formulation.  {However,}  in contrast to the experimental data a pronounced peak structure was also found in the lowest-order cumulant ratios of the net-proton fluctuations  even for small values of couplings~\cite{Bluhm:2016byc}.

{An additional complication arises from the fact that the relevant quark masses $m_q$ are finite but small. In the limit of vanishing $m_q$, there is a second-order phase transition for small and even zero $\mu_B$ belonging to the O(4) static universality class. This line of second-order phase transitions terminates in a tricritical point~\cite{Hatta:2002sj}. Recent lattice QCD studies~\cite{Ding:2019prx,Karsch:2019mbv} and measurements of higher-order cumulants of net-particle multiplicity distributions~\cite{Nonaka:2019fhk} have revealed a significant impact of the hidden O(4) criticality at larger $\sqrt{s}$ even in the case of explicit chiral symmetry breaking. Moreover, in~\cite{Hatta:2002sj} it was shown that for reasonable light quark masses the physics near CP is strongly affected by the presence of the tricritical point. In fact, only very close to CP, for distances smaller than $1.5$~MeV in $\mu_B$, criticality in line with Z(2) has been observed. Outside, in a wider region around CP, traces of the hidden tricritical point become visible as a difference in the scaling behavior of chiral and net-baryon number susceptibilities. From a phenomenological point of view it is, therefore, reasonable to attempt an implementation of the overlap of the two static universality classes relevant for the QCD phase transition, O(4) and Z(2).}

The  objective of this work is to re-examine the model assumptions introduced in Ref.~\cite{Bluhm:2016byc} to  improve  the discrepancies between the  model predictions and   STAR data on the variance of net-proton number fluctuations obtained in heavy-ion collisions.  The general {idea} is based on the universality for the critical scaling behavior of the net-quark (net-baryon) number susceptibility~\cite{Hatta:2002sj} {in the vicinity of CP that is influenced by the tricritical point}. We will show that the consistent implementation of  critical scaling   leads to a  much weaker  singularity than that   seen for the net-proton number variance in  Ref.~\cite{Bluhm:2016byc}. In the mean field approximation of the chiral effective models, the  singular part of
the net-quark number susceptibility is proportional to the chiral susceptibility multiplied by  the order parameter squared~\cite{Sasaki:2006ws,Sasaki:2007qh}. {This is correct both for O(4) and the tricritical point.} 
In line with these findings, we propose the  modification of the  model  introduced in Ref.~\cite{Bluhm:2016byc} to further  consistently  identify the influence of criticality  due to the existence of the CP on the net-proton number fluctuations {in such a way that the impact of the O(4) criticality is taken into account}.
We will quantify different cumulant  ratios of these fluctuations along  the  chemical freeze-out line obtained   in heavy-ion collisions. We will also identify systematics of the net-proton number cumulant ratios  for different locations of the CP in the $(T,\mu_B)$-plane, relative to the freeze-out line.

The theoretical tools used in our study are introduced  in Section~\ref{Sec:Theory}. {In particular in Section~\ref{Sec:TheoryRefinement} we outline our procedure for capturing the overlap between the O(4) and Z(2) criticalities phenomenologically.} In Section~\ref{Sec:Results}, we present results  on the properties of the  net-proton number fluctuations in the presence of the CP and for different assumptions on its location or the coupling strength between particles and the critical mode.   We conclude our findings in Section~\ref{Sec:Conclusions}.

\section{Modeling critical fluctuations near CP\label{Sec:Theory}}

In the following, we describe the theoretical tools to be used in order to quantify the net-proton number fluctuations in the presence of the QCD critical point. We first define the  baseline model which does not contain contributions from chiral critical mode fluctuations. Then we explain how the critical fluctuations can be coupled to   particles {near the chiral CP}. Our approach is motivated by the observation that the critical contribution to the variance of the net-proton number should obey a certain scaling behavior~\cite{Sasaki:2006ws,Sasaki:2007qh}. On the mean-field level, this idea can be extended to higher-order fluctuations. The critical mode fluctuations are obtained by using universality class arguments between QCD and the $3$-dimensional Ising spin model. The necessary mapping between the corresponding variables is also discussed.


\subsection{Baseline model}

As a  baseline model for calculating the net-proton number cumulants we employ the hadron resonance gas (HRG). In this model, the pressure of the interacting hadron gas is approximated by the sum of the partial pressures of non-interacting hadrons and their resonances~\cite{NPBM,Venugopalan:1992hy}. In the HRG model, the particle density of each particle species is given by the  ideal gas expression
\begin{equation}
\label{equ:particledensity}
n_i(T,\mu_i)=d_i\int\frac{d^3k}{(2\pi)^3}f_i^0(T,\mu_i)\,,
\end{equation}
where $d_i$ is the degeneracy factor and
\begin{equation}
\label{equ:distfunction}
f_i^0=\frac{1}{(-1)^{B_i}+e^{(E_i-\mu_i)/T}}
\end{equation}
is the thermal equilibrium distribution function. In Eq.~(\ref{equ:distfunction}), $E_i=\sqrt{k^2+m_i^2}$ and $\mu_i=B_i\mu_B+S_i\mu_S+Q_i\mu_Q$ are the energy and chemical potentials of a particle with mass $m_i$, baryon number $B_i$, strangeness $S_i$ and electric charge $Q_i$, and $\mu_X$ is the conjugate to the conserved charge number $N_X$. For a grand canonical ensemble, the average number in a constant volume $V$ is $\langle N_i\rangle=Vn_i$.

In the thermal medium the particle number fluctuates around its mean on an event-by-event basis. These fluctuations can be quantified in terms of cumulants, where the n-th order cumulant is defined as
\begin{equation}
C^i_n=VT^3\frac{\partial^{n-1} (n_i/T^3)}{\partial(\mu_i/T)^{n-1} }\bigg\vert_T \,,
\end{equation}
at constant  temperature $T$. In the following,   we consider the first four cumulants of the net-proton number $N_{p-\bar{p}}=N_p-N_{\bar{p}}$, which are given by
\begin{equation}
C_1=\langle N_{p-\bar{p}}\rangle =\langle N_p\rangle-\langle N_{\bar{p}}\rangle \,,
\label{eq:c_1_def}
\end{equation}
\begin{equation}
C_2=\langle (\Delta N_{p-\bar{p}})^2\rangle =C_2^p+C_2^{\bar{p}} \,,
\label{eq:c_2_def}
\end{equation}
\begin{equation}
C_3=\langle (\Delta N_{p-\bar{p}})^3\rangle =C_3^p-C_3^{\bar{p}} \,,
\label{eq:c_3_def}
\end{equation}
\begin{equation}
C_4=\langle (\Delta N_{p-\bar{p}})^4\rangle_c =C_4^p+C_4^{\bar{p}} \,,
\label{eq:c_4_def}
\end{equation}
where
\begin{equation}
\Delta N_{p-\bar{p}}= N_{p-\bar{p}}-\langle N_{p-\bar{p}}\rangle\,,
\end{equation}
\begin{equation}
\langle (\Delta N_{p-\bar{p}})^4\rangle_c=\langle (\Delta N_{p-\bar{p}})^4\rangle-3\langle (\Delta N_{p-\bar{p}})^2\rangle^2 \,.
\end{equation}
 The second equalities in Eqs. \eref{eq:c_2_def}-\eref{eq:c_4_def} hold only when correlations between different particle species vanish which is the case in the HRG baseline.

In the following, we will ignore the contributions stemming from resonance decays and concentrate solely on the primary proton and anti-proton numbers and their fluctuations. Since the cumulants are volume-dependent, it is useful to consider their ratios in which, ignoring volume fluctuations, the volume dependence cancels out. Consequently,
\begin{equation}
\frac{C_2}{C_1}=\frac{\sigma^2}{M}\,,\qquad \frac{C_3}{C_2}=S\sigma\,,\qquad \frac{C_4}{C_2}=\kappa\sigma^2\,,
\label{eq:ratios}
\end{equation}
where $M=C_1$ is the mean, $\sigma^2=C_2$ the variance, $\kappa=C_4/C_2^2$ the kurtosis,  and  $S=C_3/C_2^{3/2}$ the skewness. Under the assumption that experimentally measured event-by-event multiplicity fluctuations originate from a thermal source with given $T$ and $\mu_X$, one can compare the model results for the cumulant ratios with the experimental data to deduce features of the QCD phase diagram.


\subsection{Coupling to critical mode fluctuations\label{Sec:TheoryRefinement}}

Fluctuations of the chiral critical mode $\sigma$ near the QCD critical point are expected to affect various experimentally measurable quantities, including the net-proton number cumulants~\cite{Stephanov:1998dy,Stephanov:1999zu,Hatta:2003wn}. Currently, there is no general prescription of how to model the effect of critical fluctuations on observables. In~\cite{Bluhm:2016byc}, inspired by the way in which the particle mass is generated near the chiral transition in the sigma models, the critical mode fluctuations were incorporated into the HRG model by allowing the particle mass to fluctuate on an event-by-event basis around its mean value. Consequently, this  leads to fluctuations in the distribution function, $f_i= f_i^0+\delta f_i$. Here, the change of the distribution function due to critical mode fluctuations reads
\begin{equation}
\delta f_i=\frac{\partial f_i}{\partial m_i}\delta m_i=-\frac{g_i}{T}\frac{v_i^2}{\gamma_i}\delta\sigma \,,
\end{equation}
where $g_i$ is the coupling strength between $\sigma$ and the particle of type $i$, which in principle can also depend on $T$ and $\mu_X$, $v_i^2=f_i^0((-1)^{B_i}f_i^0+1)$ and $\gamma_i=E_i/m_i$.

Due to the above  modification of the distribution function, proton and anti-proton number fluctuations are no longer independent. Considering only the most singular contributions to the fluctuations (see~\cite{Bzdak:2016sxg} for a discussion of the impact of less critical contributions), one obtains the following expressions for the net-proton number cumulants influenced by the conjectured critical point,
\begin{equation}
C_2=C_2^p+C_2^{\bar{p}}+\langle(V\delta\sigma)^2\rangle(I_p-I_{\bar{p}})^2,
\label{eq:c2_model}
\end{equation}
\begin{equation}
C_3=C_3^p-C_3^{\bar{p}}-\langle(V\delta\sigma)^3\rangle(I_p-I_{\bar{p}})^3
\end{equation}
and
\begin{equation}
C_4=C_4^p+C_4^{\bar{p}}+\langle(V\delta\sigma)^4\rangle_c(I_p-I_{\bar{p}})^4,
\end{equation}
where
\begin{equation}
I_i=\frac{g_id_i}{T}\int\frac{d^3k}{(2\pi)^3}\frac{v_i^2}{\gamma_i}\,
\label{eq:I_integral}
\end{equation}
and $\langle(V\delta\sigma)^n\rangle_c$ are the critical mode cumulants.

To calculate the cumulants of the critical mode, we apply universality class arguments which state,  that close to the critical point different physical systems belonging to the same universality class,  exhibit the same critical behavior characterized by the corresponding critical exponents~\cite{Zinn-Justin}. Assuming that the QCD critical {end} point belongs to the same universality class as the $3$-dimensional Ising model, we identify the order parameter in QCD, $\sigma$, with the magnetization $M_{I}$, i.e.~the order parameter of the Ising spin model. This allows us to define the critical mode cumulants as
\begin{equation}
\langle(V\delta\sigma)^n\rangle_c=\left(\frac{T}{VH_0}\right)^{n-1}\left.\frac{\partial^{n-1}M_I}{\partial h^{n-1}}\right\vert_r,
\label{eq:m_cumulant}
\end{equation}
where $r=(T-T_c)/T_c$ and $h=H/H_0$ are the reduced temperature and magnetic field in the spin model, respectively, and the critical point is located at $r=h=0$.

In our approach, the second cumulant $C_2$ of the net-proton number receives critical contributions through the coupling of (anti-)protons to the critical mode via the first derivative of the magnetization $\partial M_I/\partial h$. In the spin model, this quantity is related to the magnetic susceptibility,  and because of universality, to the chiral susceptibility ($\chi$) in QCD. {For Z(2), the} chiral susceptibility is known to diverge {with a} stronger {amplitude but the same critical exponent as} the net-baryon number susceptibility $\chi_{B}$. {This is, however, correct only extremely close to the CP~\cite{Hatta:2002sj}. Further away, in a larger region around CP, traces of the hidden tricritical point lead to different critical exponents, making $\chi$ to diverge stronger than $\chi_B$~\cite{Hatta:2002sj,Sasaki:2006ws,Sasaki:2007qh}. Therefore,} the model in its form introduced in Ref.~\cite{Bluhm:2016byc} requires some modifications. These can be accomplished by using the following {chiral} relation between the net-baryon number and the chiral susceptibilities, which was found in the effective model calculations~\cite{Sasaki:2006ws,Sasaki:2007qh} {for O(4) and the tricritical point},
\begin{equation}
\chi_{B}\simeq\chi_{B}^{\rm{reg}}+\sigma^2\chi \,.
\label{eq:chi_mumu_fix}
\end{equation}
Here $\chi_{B}^{\rm{reg}}$ is the regular part of the net-baryon number susceptibility, whereas the second term in the above equation constitutes  the singular part,  $\chi_{B}^{\rm{sing}}$. Although this relation holds  on the mean-field level,  we will still use  it  in the present model  to capture the correct {dominant} scaling behavior near {but not exactly at} the QCD critical {end} point.

Following Eq.~(\ref{eq:c2_model}), the singular part of the second-order cumulant is  written,  as
\begin{equation}
C_2^{\rm sing}=\langle(V\delta\sigma)^2\rangle\,m_p^2\,(J_p-J_{\bar{p}})^2,
\label{eq:c2_crit_part}
\end{equation}
where
\begin{equation}
J_i=\frac{g_id_i}{T}\int\frac{d^3k}{(2\pi)^3}\frac{v_i^2}{E_i}\, .
\label{eq:J_i}
\end{equation}
In the linear-$\sigma$ models, the proton mass is related to the chiral condensate. By replacing the factor $m_p$ in Eq.~\eref{eq:c2_crit_part} by $g_p\sigma$, the critical contribution to the second-order cumulant in Eq.~\eref{eq:c2_model} has  the same form as the one in Eq.~\eref{eq:chi_mumu_fix}. We note that this replacement is only done at leading-order in the derivation, i.e. factors of $E_i$ remain unaffected. This phenomenological procedure results in   the following  modification
\begin{equation}
\gamma_i\ \rightarrow\ E_i/(g_i\sigma)\, ,
\label{eq:replacement}
\end{equation}
which is applied  also to the higher-order cumulants of the net-proton number. Consequently, we arrive at
\begin{equation}
C_2=C_2^p+C_2^{\bar{p}}+g_p^2\sigma^2\langle(V\delta\sigma)^2\rangle(J_p-J_{\bar{p}})^2 \,,
\label{eq:c2_modified}
\end{equation}
\begin{equation}
C_3=C_3^p-C_3^{\bar{p}}-g_p^3\sigma^3\langle(V\delta\sigma)^3\rangle(J_p-J_{\bar{p}})^3
\label{eq:c3_modified}
\end{equation}
and
\begin{equation}
C_4=C_4^p+C_4^{\bar{p}}+g_p^4\sigma^4\langle(V\delta\sigma)^4\rangle_c(J_p-J_{\bar{p}})^4 \,.
\label{eq:c4_modified}
\end{equation}
{By evaluating $\sigma$ and its cumulants via Eq.~\eref{eq:m_cumulant} from applying the scaling equation of state of the $3$-dimensional Ising model, see Section~\ref{sec:EoS}, we aim at capturing the crossover between the O(4) and Z(2) universality classes near the CP. As a result, $C_2$ now scales as $R^{\beta(3-\delta)}$ instead of $R^{\beta(1-\delta)}$ with the distance $R$ from the CP in the region that is dominated by the influence of the tricritical point. It should only approach the scaling $\propto R^{\beta(1-\delta)}$ extremely close to the critical end point. Nonetheless, we note that we will never consider a situation in which we are substantially close enough to see the pure Z(2)-criticality.} 

{In QCD} with finite quark masses the proper order parameter {of Z(2)} is a linear combination of scalar density, quark number density and energy density. Along this direction the grand potential exhibits zero curvature at the Z(2) CP~\cite{Fujii:2003bz}. The soft mode emerges as a space-like collective excitation in the scalar spectral function~\cite{Fujii:2004jt}. A massive $\sigma$-mode appears in the time-like sector and decouples from the critical fluctuations near the CP. {In our study, the $\sigma$ in Eq.~\eref{eq:chi_mumu_fix} must be interpreted as an effective soft mode for the overlap between the O(4) and Z(2) criticalities even in the presence of finite quark masses.} In the following, we use $g$ instead of $g_p$ and apply  Eqs. \eref{eq:c_1_def} and \eref{eq:c2_modified}-\eref{eq:c4_modified} together with Eq.~\eref{eq:J_i} to calculate  the net-proton number cumulant ratios.


\subsection{Magnetic equation of state\label{sec:EoS}}

For the magnetic equation of state   we use a parametric representation,  ~\cite{Guida:1996ep}
\begin{equation}
M_I=M_0R^\beta \f \, ,
\label{eq:eos_1}
\end{equation}
which is strictly speaking  valid only in the scaling region close to the critical point.  Here $R$ and $\f$ are auxiliary variables which depend on $r$ and $h$, from which $R$ measures the distance from the critical point. They are determined by solving the following equations,
\begin{eqnarray}
r&=&R(1-\f^2), \label{eq:eos_2} \\
h&=& R^{\beta\delta}w(\f), \label{eq:eos_3}
\end{eqnarray}
where $\beta$ and $\delta$ are critical exponents and
\begin{equation}
w(\f)=c\f(1+a\f^2+b\f^4) \label{eq:eos_4}
\end{equation}
is an odd polynomial in $\f$.  The parameters entering Eqs.~\eref{eq:eos_1}-\eref{eq:eos_4} are determined numerically  by the Monte-Carlo simulations   or by other theoretical tools such as the $\epsilon$-expansion or functional renormalization group methods. In~\cite{Guida:1996ep}, renormalization group and field theoretical methods were used to determine the critical exponents and coefficients of $w(\f)$,  reading,   $\beta=0.325$ and $\delta=4.8169$, and $a=-0.76145$, $b=0.00773$ and $c=1$. For the normalization constants $H_0$ and $M_0$ in Eqs.~(\ref{eq:m_cumulant}) and~(\ref{eq:eos_1}) we follow~\cite{Bluhm:2016byc} and set exemplarily,  $M_0=5.52\times 10^{-2}\,\rm{GeV}$ and $H_0=3.44\times 10^{-4}\,\rm{GeV}^3$. Differentiation of Eqs.~\eref{eq:eos_1}-\eref{eq:eos_4} with respect to $h$ as defined in Eq.~\eref{eq:m_cumulant} allows us   to determine the cumulants of the critical mode. These expressions are summarized in the appendix  of Ref. \cite{Bluhm:2016byc},  and will not be repeated here.

The parametrization of the magnetic equation of state used in this work provides an accurate description of the order parameter close to the critical point. In a simpler, linear parametric representation the coefficients and critical exponents read,  $a=-2/3$, $b=0$ and $c=3$, $\beta=1/3$ and $\delta=5$. In this representation, the cumulants of the critical mode are considerably simpler and can be found in~\cite{Mukherjee:2015swa,Stephanov:2011pb}. We note,  that the numerical results presented in~\cite{Bluhm:2016byc} were obtained using this simpler representation. Qualitatively, features of the results obtained in either of these parametrizations are very similar, however the  non-monotonic structures due to the critical point are quantitatively more pronounced in the parametrization which we employ in this study.


\subsection{Mapping between spin model and QCD\label{sec:Mapping}}

\begin{figure}[t]
\centering
  \includegraphics[width=0.6\linewidth]{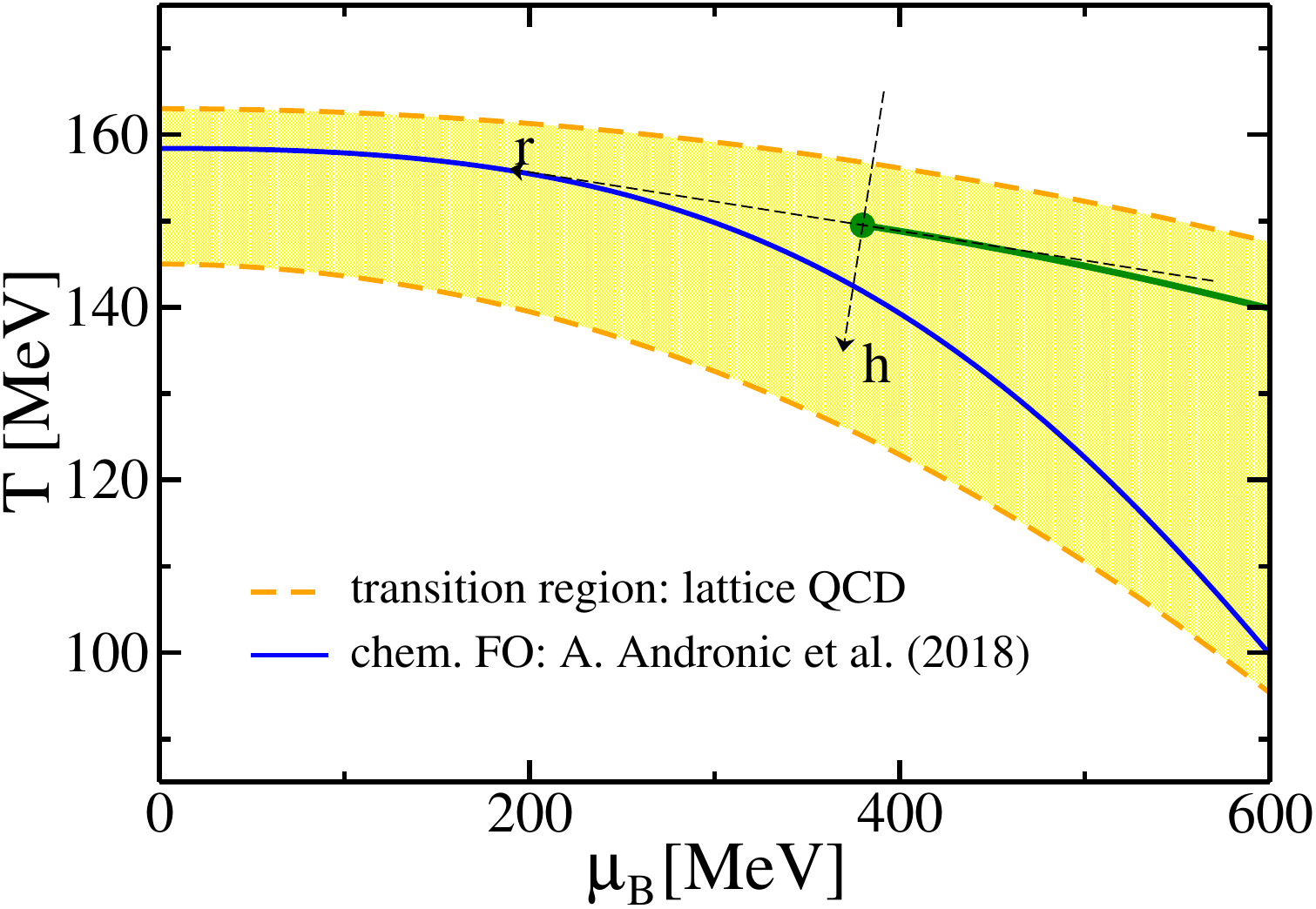}
\caption{(Color online) The model setup for the location of the CP.  The filled band between the two dashed curves shows lattice QCD results for the location of the chiral crossover  transition  obtained by solving Eq.~\eref{eq:tc_param} at leading-order for $\kappa_c=0.007$ (upper curve, $T_{c,0}=0.163\,$GeV) and $\kappa_c=0.02$ (lower curve, $T_{c,0}=0.145\,$GeV). The green dot shows the critical point with the attached spin model coordinate system (see the main text for details of the mapping) and the first-order phase transition line for larger $\mu_B$. The solid blue line shows the chemical freeze-out curve from~\cite{NPBM}.}
\label{fig:setup}
\end{figure}

In order to utilize the universality class argument, one needs a mapping between the reduced temperature $r$ and magnetic field $h$ in the spin model and the QCD temperature $T$ and baryon-chemical potential $\mu_B$. Such a mapping is non-universal. Moreover, it is sensitive to the model assumptions. One of the frequently used models is a linear mapping between the spin model and QCD phase diagrams~\cite{Mukherjee:2015swa,Nonaka:2004pg}. There, the main assumptions in  the mapping are: (i) The conjectured QCD critical point at $(\mu_{cp},T_{cp})$ is located at $r=h=0$ in the Ising model coordinate system, and (ii) the $r$ axis is tangential at the critical point to the first-order phase transition line in QCD, where the positive $r$ direction points towards the QCD crossover region. The orientation of the $h$ axis is not well constrained. In this work we assume that this axis is perpendicular to the $r$ axis and its positive direction points towards the hadronic phase of QCD (see Fig.~\ref{fig:setup}).

To obtain $(r,h)$ corresponding to a given $(\mu_B,T)$ pair it is convenient to introduce an auxiliary coordinate system $(\tilde{r},\tilde{h})$, originating at the QCD critical point and oriented such that the $\tilde{r}$ axis is parallel to the $\mu_B$ axis. Then the mapping is defined as
\begin{equation}
\tilde{r}=\frac{\mu_B-\mu_{cp}}{\Delta \mu_{cp}},
\end{equation}
and
\begin{equation}
\tilde{h}=\frac{T-T_{cp}}{\Delta T_{cp}},
\end{equation}
where $\Delta T_{cp}$ and $\Delta \mu_{cp}$ are parameters which are connected to the size of the critical region. Following~\cite{Bluhm:2016byc} we set $\Delta T_{cp}=0.02\,$GeV and $\Delta \mu_{cp}=0.42\,$GeV. The corresponding point in the spin model coordinate system is obtained by rotation of the auxiliary coordinate system, where the angle is determined by the slope of the first-order phase transition line of QCD at the critical point.

The exact location of the QCD critical point and the slope of the first-order phase transition line are not known. Input provided by lattice QCD calculations may be used to constrain these parameters. The crossover line can be para\-me\-tri\-zed as
\begin{equation}
T_c(\mu_B)=T_{c,0}\left[1-\kappa_c\left(\frac{\mu_B}{T_c(\mu_B)}\right)^2+...\right],
\label{eq:tc_param}
\end{equation}
where $T_{c,0}=(0.145\, \dots\, 0.163)\,$GeV is the critical temperature at vanishing chemical potential \cite{L3,L4} and $\kappa_c\,\simeq\,0.007\, \dots\, 0.059$ is the chiral crossover curvature \cite{Kaczmarek:2011zz,Endrodi:2011gv,Bonati:2015bha,Cea:2015cya}. Then, for given $\mu_{cp}$, $T_{c,0}$ and $\kappa_c$ the temperature $T_{cp}$ and the slope of the first-order phase transition line at $(\mu_{cp},T_{cp})$ are obtained from Eq.~\eref{eq:tc_param}.

Finally, to make contact between our model calculations and the experimental data on net-proton fluctuations we calculate the net-proton number cumulants at chemical freeze-out. In this work we use the chemical freeze-out conditions which were determined by analyzing the measured hadron yields~\cite{Abelev:2013vea,Abelev:2013xaa,ABELEV:2013zaa,Abelev:2014uua,Adam:2015yta,Adam:2015vda}. The solid blue line in Fig.~\ref{fig:setup} indicates the  recent parametrization from Ref.~\cite{NPBM}.


\section{Numerical results on the net-proton number fluctuations\label{Sec:Results}}

In the following we discuss the  numerical results  on different fluctuations of the net-proton number obtained in  the above  phenomenological   approach that correctly embeds the expected scaling behavior of the net-proton variance~\cite{Sasaki:2006ws,Sasaki:2007qh}. To link  our results to STAR  data  we first  discuss how the experimentally applied kinematic acceptance cuts can be included into the framework. Then, the impact of our phenomenological modification introduced in Eq.~\eref{eq:replacement}  on fluctuation observables  is studied. This includes also a discussion of the effects of modifying the coupling strength $g$ and the proximity of the thermal conditions at which the cumulant ratios are evaluated near  the QCD critical point.


\subsection{Kinematic cuts}

In the fluctuation measurements, the investigated phase-space coverage is limited by the detector design and specific demands from the experimental analysis, like e.g.  to optimize efficiency. Since the observables depend  on the implemented  kinematic acceptance, (see e.g. refs.~\cite{Adamczyk:2013dal} and~\cite{Luo:2015ewa,Luo:2015doi,Thader:2016gpa,Morita}) it is important  to incorporate the experimental cuts into our theoretical framework. Following~\cite{Garg:2013ata}, we include restrictions in kinematic rapidity $y$, transverse momentum $k_T$ and azimuthal angle $\phi$ by replacing
\begin{equation}
\int d^3k\ \longrightarrow\ \int k_T\sqrt{k_T^2+m_i^2}\cosh y \,dk_T\,dy\,d\phi,
\end{equation}
and $E_i\to \sqrt{k_T^2+m_i^2}\cosh y$ in the momentum-integrals. In line with~\cite{Thader:2016gpa}, we consider the following  phase-space integrations: $-0.5\leq y \leq 0.5$, $0\leq \phi \leq 2\pi$ and $0.4\,\rm{GeV/c}\leq k_T \leq 2\,\rm{GeV/c}$. We note,  that this procedure cannot account for scattering of particles in and out of the acceptance window during the late stage  evolution of a medium created in heavy-ion collisions.


\subsection{Net-proton number cumulant ratios}

The contributions of critical fluctuations to the net-proton number cumulants  are sensitive to the value of the coupling $g$ between (anti-)protons and the critical mode. This value may, in principle, depend on $T$ and $\mu_X$, i.e.~on the position in the QCD phase diagram where the cumulant ratios are evaluated. In fact,  in the quark-meson~\cite{AbuShady:2015ab} and NJL \cite{Hatsuda:1987ab} models, the  meson-nucleon couplings are found to decrease both with increasing $T$ and/or $\mu_B$. In the following, we consider fixed values for $g$ along the chemical freeze-out curve depicted in Fig.~\ref{fig:setup}, i.e.  independent of the beam energy $\sqrt{s}$. Typical values for $g$ may be inferred from various effective model calculations. In the linear sigma model  this parameter can be related in the ground state to the pion decay constant, $g\simeq m_p/f_\pi\simeq 10$~\cite{Stephanov:2008qz,Hatta:2003wn}. Similar values can be found in non-linear chiral models~\cite{Dexheimer:2008ax} describing QCD matter in neutron stars. On the other hand, based on different  quark-meson models the value of  $g\simeq 3-7$ for the nucleon-meson couplings is  well conceivable~\cite{Downum:2006re}. To highlight the features of our model results we will use $g$ in the range between $3$ and $5$. 

\begin{figure}[t]
\centering
\minipage{0.5\linewidth}
\includegraphics[width=\linewidth]{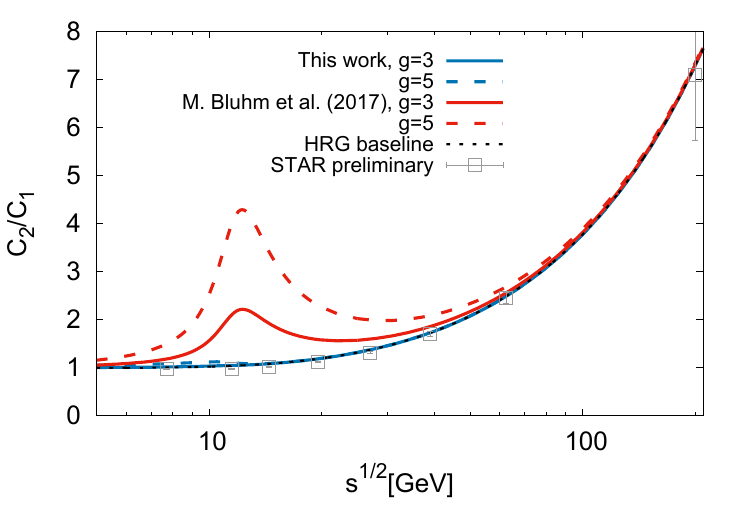}
\endminipage
\minipage{0.5\linewidth}
\includegraphics[width=\linewidth]{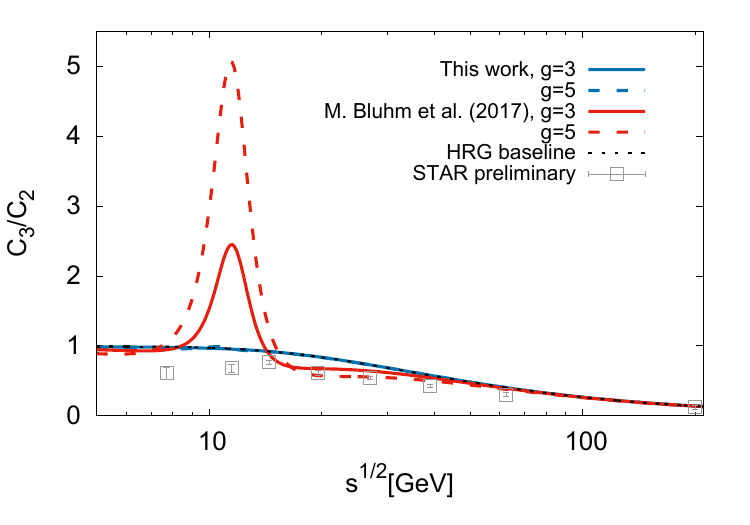}
\endminipage\\
\minipage{0.5\linewidth}
\includegraphics[width=\linewidth]{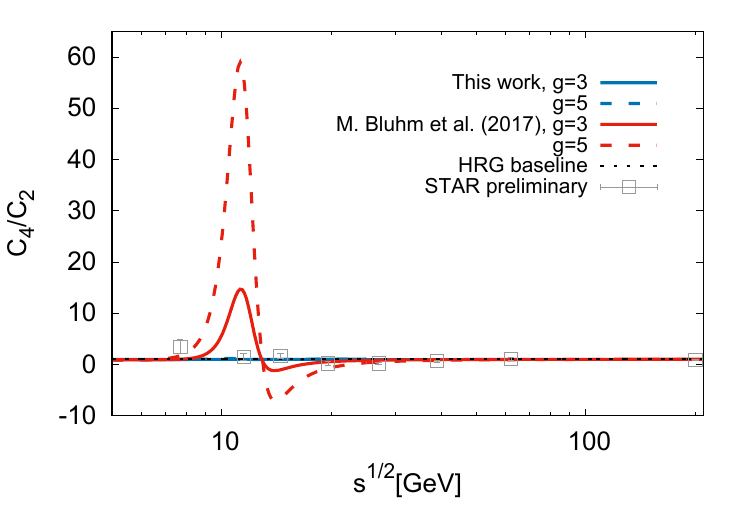}
\endminipage
\caption{(Color online) Net-proton number cumulant ratios from  Eq.~\eref{eq:ratios} calculated following  Ref.~\cite{Bluhm:2016byc} for $g=3$ and $5$ (red solid and dashed lines, respectively) in comparison with present model results (blue solid and dashed lines, respectively). We note that we modified the set-up compared to Ref.~\cite{Bluhm:2016byc}, i.e. we changed the orientation of the $h$-axis, neglected resonance decays, considered different freeze-out conditions and applied a different parametrization of the magnetic equation of state. For comparison, we also show the preliminary STAR data on the net-proton number fluctuations~\cite{Thader:2016gpa} (squares, where the error bars contain both statistical and systematic errors). Also shown are results for the non-critical baseline (black dotted lines).}
\label{fig:imp_vs_orig}
\end{figure}

In the previous work~\cite{Bluhm:2016byc}, the critical point was exemplarily located at $\mu_{cp}=0.39$~GeV and $T_{cp}=0.149$~GeV (see Fig.~\ref{fig:setup}). There, even for rather small values of $g\simeq 3$, the cumulant ratio $C_2/C_1$ exhibited a clear peak structure compared to the non-critical baseline and in contrast to the STAR data~\cite{Adamczyk:2013dal,Luo:2015ewa,Luo:2015doi,Thader:2016gpa}. This is illustrated in Fig.~\ref{fig:imp_vs_orig}. 

{The effect of the modified scaling,  discussed in } Sec.~\ref{Sec:TheoryRefinement},  is the substantial reduction of the critical contribution to $C_2$ in Eq.~\eref{eq:c2_modified}  implying the  disappearance of the maximum in $C_2/C_1$ even for  large values of $g\simeq 5$, as seen in Fig.~\ref{fig:imp_vs_orig}. Within error bars, the model results  are in agreement with  data of the STAR Collaboration~\cite{Thader:2016gpa} on the $C_2/C_1$ ratio. We note, that in order to see a similar maximum in  $C_2/C_1$  in  this   model calculations,  as found   in~\cite{Bluhm:2016byc}  for a given location of the critical point,  a significantly larger value of $g$,  outside the expected  range discussed  above,  would be necessary. {Furthermore, the differences between the} $C_2/C_1$ ratios shown in Fig.~\ref{fig:imp_vs_orig} are independent of the particular choice for the phenomenological freeze-out conditions. Indeed, in these    studies we have adopted the freeze-out line from Ref.  \cite{NPBM}, whereas in \cite{Bluhm:2016byc}  the $C_2/C_1$ ratio was calculated along the freeze-out line from Refs. \cite{Alba:2014eba,Bluhm:2014vua}.

The higher-order cumulant ratios of the net-proton number fluctuations  are also shown in Fig.~\ref{fig:imp_vs_orig}. The model introduced in~\cite{Bluhm:2016byc} exhibits clearly pronounced non-monotonic structures of higher-order cumulant ratios for $g=3$ and $5$. We note, that while in the model~\cite{Bluhm:2016byc} the ratios $C_2/C_1$ and $C_4/C_2$ are independent of the choice made for the orientation of the $h$-axis, the behavior of $C_3/C_2$ is sensitive to this choice. In~\cite{Bluhm:2016byc}, the positive direction of the $h$-axis was defined to point upward towards larger values of $T$. In our work, we choose the opposite direction as discussed in Sec.~\ref{sec:Mapping} such that $g\sigma$ is positive along the chemical freeze-out curve.

As seen in Fig.~\ref{fig:imp_vs_orig}, in the present model calculations, the non-monotonic structures of the higher-order cumulant ratios become strongly suppressed even for $g=5$. In fact, for the considered setup, the model results show rather small deviations from the non-critical baseline. Moreover, in contrast to~\cite{Bluhm:2016byc}, the behavior of $C_3/C_2$ in the present model (blue lines in Fig.~\ref{fig:imp_vs_orig}) does not depend on the orientation of $h$. This is because the combined $\f$-dependence in $C_3$ of the critical mode fluctuations $\langle(V\delta\sigma)^3\rangle$ and the factor $\sigma^3$ is even, see Eq.~\eref{eq:c3_modified}. Thus, although $h$ is an odd function in $\f$ as seen from Eqs.~\eref{eq:eos_3} and~\eref{eq:eos_4}, a re-orientation of the $h$-axis would have no effect.

The substantial reduction of the critical signal  in the net-proton number cumulant ratios seen in  our  model  results is  a consequence of (i) the reduced scaling of the critical contributions to  $C_{n=2,3,4}$ in Eqs.~\eref{eq:c2_modified}-\eref{eq:c4_modified} and (ii) the magnitude  of the factors $(g\sigma)^n$. As a result of the  phenomenological implementation in  Eq.~\eref{eq:replacement}, the scaling of $C_n$ at the CP is weakened by an additional $n\beta$  factor in the critical exponents.
{ Moreover, the factor $g\sigma$ differs from the vacuum proton mass $m_p$ in Eq.~\eref{eq:c2_crit_part} as  employed in~\cite{Bluhm:2016byc}. In the present calculations,   $g\sigma$ is of the order of $0.2-0.3$~GeV for most $\sqrt{s}$. Its actual values depend on the parameters in the magnetic equation of state, most notably  on the value of $M_0$ (see Sec.~\ref{sec:EoS}), and the mapping between spin model and QCD (see Sec.~\ref{sec:Mapping}). These values receive support from recent works on the origin of the baryon masses in both lattice QCD~\cite{Aarts:2015mma,Aarts:2017rrl} and effective models based on parity doubling~\cite{Sasaki:2017glk,Marczenko:2017huu}. There, the baryon masses are found to be given to a large extent by $\sigma$-independent contributions.}

\begin{table}[t]
\caption{Considered locations  of the QCD critical point in the ($\mu_B,T$)-plane.   The parameters $T_{c,0}$ and $\kappa_c$ of the crossover (pseudo-critical) line Eq.~\eref{eq:tc_param} needed to determine $T_{cp}$ and the slope of the first-order phase transition line at the critical point for a given $\mu_{cp}$ are also listed. The locations of these critical points in the QCD phase diagram are shown in Fig.~\ref{fig:cp_setup}.}
\label{tab:cp_setup}
\begin{indented}
\lineup
\item[]\begin{tabular}{@{}*{5}{l}}
\br
$CP_i$ & $\mu_{cp}\,$[GeV] & $T_{cp}\,$[GeV] & $T_{c,0}\,$[GeV] & $\kappa_c$\cr
\mr
1      &  0.390           &  0.149         &  0.156          & 0.007      \cr
2      &  0.420           &  0.141         &  0.155          & 0.010      \cr
3      &  0.450           &  0.134         &  0.155          & 0.012      \cr
\br
\end{tabular}
\end{indented}
\end{table}

\begin{figure}[t]
\centering
  \includegraphics[width=0.6\linewidth]{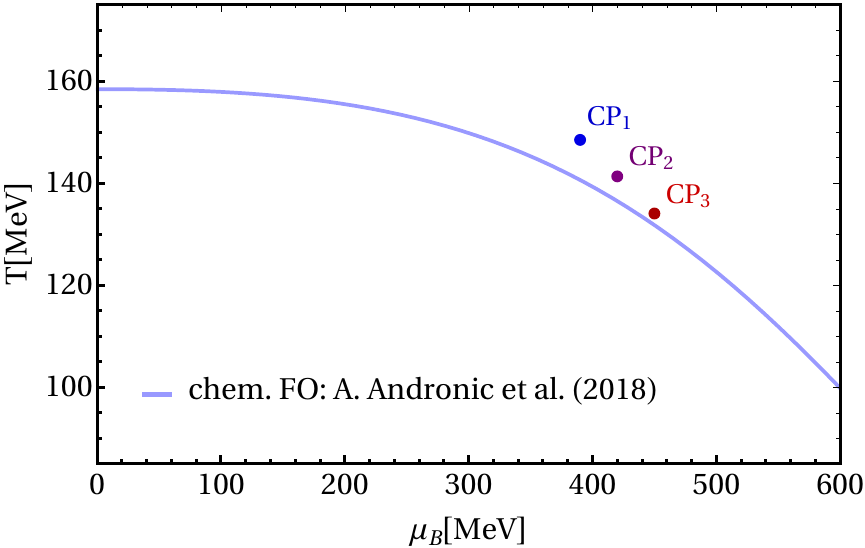}
\caption{(Color online) Locations of the QCD critical points from Tab.~\ref{tab:cp_setup} plotted together with the chemical freeze-out curve~\cite{NPBM} used in this work.}
\label{fig:cp_setup}
\end{figure}

The properties of different cumulant ratios shown  in  Fig.~\ref{fig:imp_vs_orig}  can lead to the conclusion, that the monotonic beam energy dependence seen experimentally in $C_2/C_1$  together with the non-monotonicity in the higher-order cumulant ratios cannot be  explained simultaneously by a model that (i) includes critical mode fluctuations through the coupling of $\sigma$ with the particles and (ii) obeys the connection between the chiral and net-baryon number susceptibilities observed in effective models~\cite{Sasaki:2006ws,Sasaki:2007qh}.
 However, one  notes that the model  results depend not only on the values of the coupling $g$ but  also on  the non-universal details of the  mapping between QCD and the spin model discussed in Sec.~\ref{sec:Mapping}. One of them   is the unknown distance of the QCD critical point from the chemical  freeze-out conditions at which the fluctuations are determined. To study this effect we keep the chemical freeze-out conditions fixed but vary the location of the critical point in the QCD phase diagram as summarized in Tab.~\ref{tab:cp_setup} and depicted  in Fig.~\ref{fig:cp_setup}.

\begin{figure}[t]
\centering
\minipage{0.5\linewidth}
  \includegraphics[width=\linewidth]{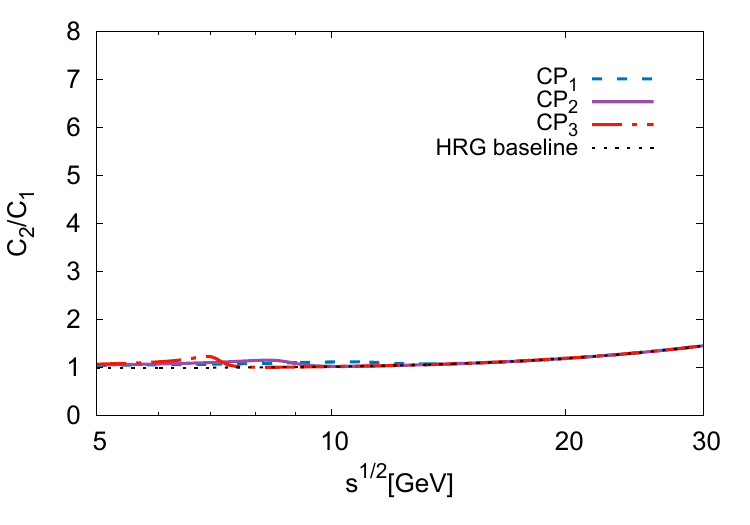}
\endminipage
\minipage{0.5\linewidth}
  \includegraphics[width=\linewidth]{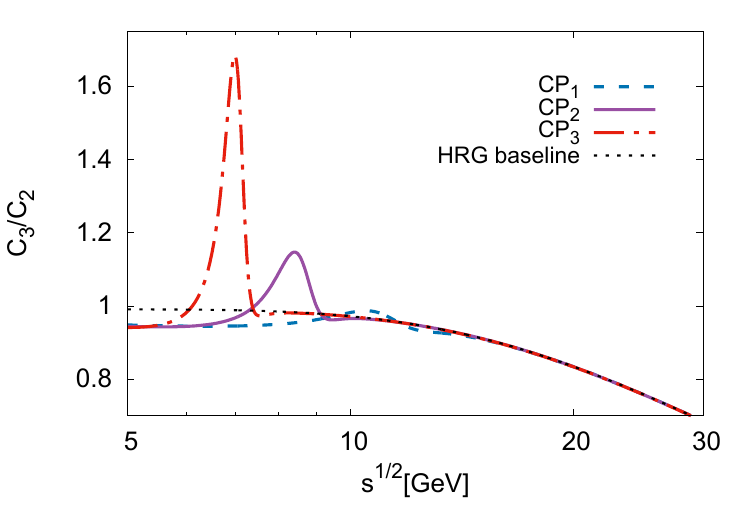}
\endminipage\\
\minipage{0.5\linewidth}
  \includegraphics[width=\linewidth]{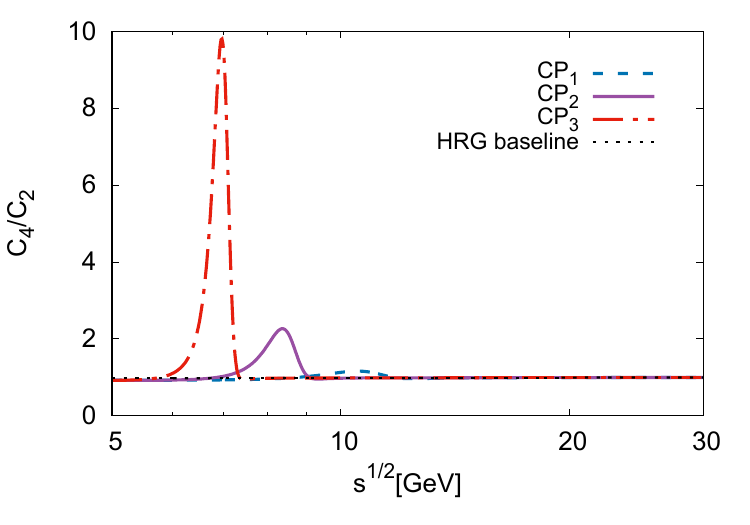}
\endminipage
\caption{(Color online) Net-proton number cumulant ratios from  Eq.~\eref{eq:ratios} calculated    in the present   model for fixed $g=5$ and for different locations of the QCD critical point as listed in Tab. \ref{tab:cp_setup}.
}
\label{fig:cp_g_5}
\end{figure}

In Fig.~\ref{fig:cp_g_5} we show the net-proton number fluctuations  along the phenomenological freeze-out line at fixed value of the coupling $g=5$ and assuming different locations of the CP.  As evident from this figure, moving the critical point closer to the chemical freeze-out curve leads to an increase of non-monotonic structures in the net-proton number cumulant ratios. While deviations from the non-critical baseline (black dotted lines) remain moderately weak in $C_2/C_1$, they become more pronounced with  increasing order of the fluctuations in  $C_n/C_2$ ratios. {We note that even for CP$_3$ we remain always at a distance of about or larger than $3$~MeV in $\mu_B$ along the chemical freeze-out line.}

\begin{figure}[t]
\centering
\minipage{0.5\linewidth}
  \includegraphics[width=\linewidth]{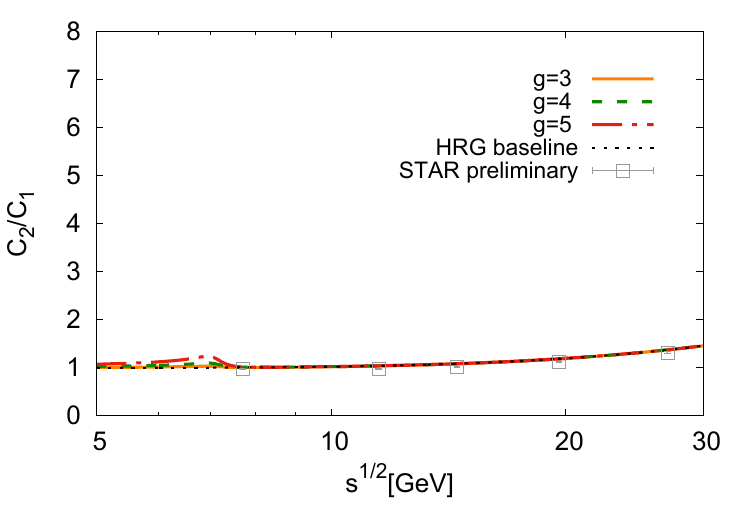}
\endminipage
\minipage{0.5\linewidth}
  \includegraphics[width=\linewidth]{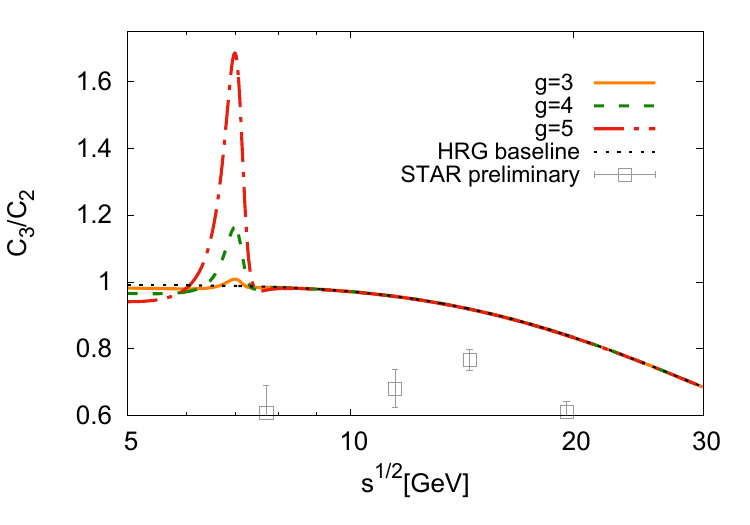}
\endminipage\\
\minipage{0.5\linewidth}
  \includegraphics[width=\linewidth]{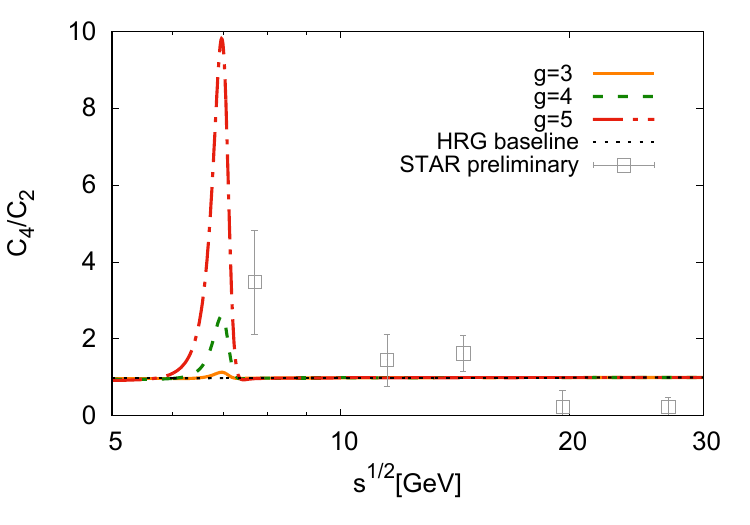}
\endminipage
\caption{(Color online) Net-proton number cumulant ratios from  Eq.~\eref{eq:ratios} calculated with the  CP$_3$-setup for the location of the CP (see Tab. \ref{tab:cp_setup}) and for different values of the coupling $g=$ 3, 4 and 5 (orange solid, green long-dashed and red dash-dotted lines, respectively). The preliminary STAR data for the net-proton number fluctuations~\cite{Thader:2016gpa} (squares, where the error bars contain both statistical and systematic errors) are shown for comparison. The non-critical baseline model results are shown by black dotted lines.}
\label{fig:cp_3_g_comparison}
\end{figure}

In Fig.~\ref{fig:cp_3_g_comparison} we show the influence of the critical point located  at the closest distance CP$_3$ on the  energy dependence of the net-proton number fluctuations 
 for different values of the coupling $g$.  As can be seen,  all cumulant ratios depend strongly on the actual value of $g$. This behavior is expected from Eqs.~\eref{eq:c2_modified}-\eref{eq:c4_modified} and Eq.~\eref{eq:J_i} since  cumulants $C_n$ scale as $g^{2n}$.
This differs from the  model introduced in
\cite{Bluhm:2016byc} where, as seen from  Eqs.~\eref{eq:c2_model}-\eref{eq:I_integral},  they scale only as $g^n$.
When compared to the STAR data, the  results for $C_2/C_1$ and $C_4/C_2$ shown in  Fig.~\ref{fig:cp_3_g_comparison} are in qualitative  agreement with  data, given the uncertainties in the model assumptions. {However, the $C_3/C_2$ ratio also  increases  beyond the non-critical baseline  towards the lower beam energies in contrast to the  STAR data. }

From the results shown in Figs.~\ref{fig:imp_vs_orig},~\ref{fig:cp_g_5} and~\ref{fig:cp_3_g_comparison} it is clear that for small couplings $g\simeq 3$ deviations from the non-critical baseline are negligible in all cumulant ratios irrespective of the studied location of the critical point. By increasing $g$, non-monotonic structures in the $\sqrt s$-dependence of the net-proton cumulant ratios  develop and are sensitive to the relative distance between the critical point and the chemical freeze-out curve. This behavior is also stronger in the higher-order cumulant ratios.

It is therefore conceivable  that,  by an  appropriate choice of the location of the CP and the model parameters, it is possible to describe the energy dependence of {some}  ratios of the net-proton number fluctuations as  seen in the preliminary STAR data.  Consequently,  the rather strong increase of the
$C_4/C_2$ ratio  beyond the HRG baseline   and the  smooth dependence  of $C_2/C_1$ observed in heavy-ion collisions at  energies $\sqrt s <20$ GeV   could be due to the contribution from the CP located near the chemical freeze-out line. {However, in this case the $C_3/C_2$ ratio should also exceed the non-critical baseline, which is not seen experimentally.}

{Thus, based on the presented equilibrium model results one concludes, that the energy dependence of the $C_4/C_2$, $C_3/C_2$ and $C_2/C_1$ ratios observed in heavy ion collisions at $\sqrt s <20$~GeV by the STAR Collaboration does not follow the systematics expected from the contribution of the CP to the net-proton number fluctuations alone. This conclusion is consistent with the previous analysis of different fluctuation observables based on lattice QCD and PNJL model results~\cite{T4,T5}. We note, however, that the above statement requires further theoretical and empirical justifications due to current uncertainties in the model assumptions and the experimental data. It remains to be seen, for example, whether non-equilibrium effects can, in fact, push the $C_3/C_2$ ratio below the non-critical HRG model baseline. Moreover, the fireball evolution or late hadronic stage processes such as resonance decays or isospin randomization, which have not been included in our study, may influence the theoretical results.}


\section{Conclusions\label{Sec:Conclusions}}

We have studied the influence of the QCD critical {end} point (CP)  on the properties of the n-th order  cumulants ($C_n$)  of the net-proton number and their ratios. The results were addressed in the context of the recent data from the STAR Collaboration on the energy dependence of the net-proton number fluctuations in Au-Au collisions  obtained wi\-thin the Beam
Energy Scan  program  at the Relativistic Heavy Ion Collider (RHIC).

To calculate the net-proton number cumulants  we have proposed  a phenomenological model where  non-critical fluctuations are  obtained from the hadron resonance gas (HRG)  statistical operator, which is known to describe data on  particle yields in heavy-ion collisions and the  lowest-order fluctuation observables from lattice QCD. For simplicity,  the baryonic  sector of the HRG   was  approximated  by  contributions from     primary  protons and anti-protons.
To quantify  the role     of  the chiral criticality due to the CP, the phenomenological model was introduced  to describe the non-analytic part of the statistical operator in which the fluctuations of the chiral critical mode  $\sigma$ are   coupled to the  (anti-)protons.  This was achieved   by linking their   masses to the $\sigma$ mode, as suggested by    different chiral models. 
Consequently, the (anti-)proton  mass and its momentum distribution function
fluctuate on an event-by-event basis around its mean or equilibrium  value, respectively.

The critical mode fluctuations were determined by applying universality class arguments between QCD and the $3$-dimensional Ising spin model. We have extended  the model introduced in Ref.~\cite{Bluhm:2016byc} by accounting for the critical scaling behavior of the net-baryon variance $\chi_B$ suggested by effective chiral models. There $\chi_B$ is linked to the product of the chiral susceptibility and the chiral order parameter squared. {In this way, we have phenomenologically embedded the overlap between a hidden O(4) and the Z(2) criticality, which is theoretically expected in the vicinity of CP up to a tiny region in $\mu_B$~\cite{Hatta:2002sj}, into the model.}

We have found a substantial reduction of the critical mode contributions to the net-proton number fluctuations compared to the results in Ref.~\cite{Bluhm:2016byc}. This is a consequence of the reduced critical scaling imposed by respecting the proper scaling relation between net-baryon number and chiral susceptibilities and the size of the proton mass modification due to the coupling to the $\sigma$ mode. This brings our results for different n-th order cumulants $(C_n)$ of the net-proton number, calculated along the phenomenological chemical freeze-out line, closer to the experimental observations made by the STAR Collaboration for the energy dependence of the cumulant ratios in heavy-ion collisions at RHIC. 
{In particular, with an appropriate choice of the model parameters and the location of the CP relative to the chemical freeze-out line, the model can reproduce the smooth energy dependence of $C_2/C_1$ and the increase and non-monotonic variation of $C_4/C_2$ towards lower beam energies, as is observed by the STAR Collaboration.
However, the decrease of the $C_3/C_2$ ratio towards lower beam energies seen in the STAR data is inconsistent with the systematics expected in the present model from the contribution of the CP, which would predict an access of this ratio beyond the non-critical baseline. Thus, our conclusion is that it is rather unlikely that the properties observed in the low energy behavior of different ratios of the net-proton number cumulants in heavy-ion collisions are due to the existence of the critical point near the phenomenological chemical freeze-out line alone.} 


\ack{
The work of M.~Szymański and M.~Bluhm is funded in parts by the European Union's Horizon 2020 research and innovation program under the Marie Sk\l{}odowska Curie grant agreement No 665778 via the Polish National Science Center (NCN) under grant Polonez UMO-2016/21/P/ST2/04035. M.~Bluhm acknowledges also the partial support by the program “Etoiles montantes en Pays de la Loire 2017”. This work is partly supported by the Polish National Science Center (NCN) under Maestro grant no. DEC-2013/10/A/ST2/00106 and Opus grant no. 2018/31/B/ST2/01663. K.~Redlich also acknowledges partial support of the Polish Ministry of Science and Higher Education.
We thank F.~Geurts and J.~Th\"ader for providing the preliminary STAR data~\cite{Thader:2016gpa} shown in Figures~\ref{fig:imp_vs_orig} and~\ref{fig:cp_3_g_comparison}. We also thank Bengt Friman and Marlene Nahrgang  for interesting discussions and comments.}


\end{document}